\documentclass[11pt,a4paper]{article} 
\usepackage{jheppub}

\usepackage{etoolbox}
\makeatletter
    \patchcmd{\maketitle}{\@fpheader}{}{}{}
    \makeatother

\usepackage{quotes}
    
\usepackage{tikz}

\usepackage{booktabs}

\usepackage{slashed}
\usepackage{amsmath}

\usepackage{float}

\usepackage{pgfplots} 
\newcommand{\scale}{0.6cm}
\pgfmathsetmacro{\range}{4}
\pgfmathsetmacro{\gridgreylevel}{30}
\pgfmathsetmacro{\margin}{0.25}
\newcommand*{\crossdiagramlayout}[1]{%
    \foreach \i in {1,...,\range} {
        \draw[very thin,black!\gridgreylevel] (-\margin,\i) node[left,black] {$\i$} -- ({\range+\margin},\i);
    }
    \foreach \i in {1,...,\range} {
        \draw[ultra thin,black!\gridgreylevel] (\i,-\margin) node[below,black] {$\i$} -- (\i,{\range+\margin});
    }

    \draw[-latex,black] (-\margin,0) node[left] {$0$} -- ([xshift=4pt]{\range+\margin},0) node[below,xshift=2pt] {$p$};
    \draw[-latex,black] (0,-\margin) node[below] {$0$} -- ([yshift=4pt]0,{\range+\margin}) node [left,yshift=2pt] {$q$};

    \node[above] at ({\range/2},{\range+\margin}) {$r=#1$};
}

\newcommand{\beq}{\begin{equation}}
\newcommand{\eeq}{\end{equation}}

\newcommand{\beqa}{\begin{eqnarray}}
\newcommand{\eeqa}{\end{eqnarray}}

\newcommand{\by}{\begin{eqnarray}}
\newcommand{\ey}{\end{eqnarray}}

\newcommand{\half}{\frac{1}{2}}

\newcommand{\mN}{\mathcal N}

\newcommand\fverb{\setbox\fverbbox=\hbox\bgroup\verb}
\newcommand\fverbdo{\egroup\medskip\noindent%
            \fbox{\unhbox\fverbbox}\ }
\newcommand\fverbit{\egroup\item[\fbox{\unhbox\fverbbox}]}
\newbox\fverbbox

\newcommand{\nablaslash}{\not{\hbox{\kern-3pt $\nabla$}}}

\title{Squashed 7-spheres, octonions and the swampland}

 \author{Bengt E.W.~Nilsson}
    \affiliation{Department of Physics, \\Chalmers University of Technology, \\SE-41296 G\"oteborg, Sweden}

\abstract{ The entire eigenvalue spectrum of the  
operators on the squashed  $S^7$ that appear in the Freund-Rubin compactification of eleven-dimensional supergravity was recently 
derived in  \cite{Nilsson:2018lof, Ekhammar:2021gsg, Nilsson:2023ctq, Karlsson:2023dnl}. Here we give a brief account of this work which started
with  \cite{Nilsson:2018lof}  where the complete spectrum of irreducible isometry  representations of the fields in $AdS_4$ was derived 
for the squashed $S^7$ compactification.   
The operator  spectra determine the mass spectrum of the fields in $AdS_4$ and are important for the corresponding  $\mN =1$ 
supermultiplet structure which appears in two versions depending on the choice of boundary conditions. By an orientation-flip on the squashed $S^7$
we can also determine the spectrum of the corresponding non-supersymmetric theory, and, e.g., its spectrum of marginal 
operators on the boundary of $AdS_4$ which may
have some relevance for the $AdS$ stability conjecture in the swampland program. 
The role of singletons is discussed and a possible new Higgsing phenomenon turning them into bulk fields is suggested. 
Details are here given primarily for 2-forms and comments are  made on  the key role of $G_2$  and octonions 
 for  the structure of the operator equations and mode functions on the squashed  $S^7$. Some important  features of these improved methods were 
 obtained  in Joel Karlsson's 2021 MSc thesis  \cite{Karlsson:2021oxd}. 
 This is an extended version of the author's contribution to the proceedings of the conference  ISQS28, Prague,  
 Czech Republic, July 1 - 5, 2024.}

\toccontinuoustrue
\begin{document}
\maketitle


\section{Introduction}
Compactification of D=11 supergravity on the squashed $S^7$ dates back to the first half of the 1980s \cite{Awada:1982pk, Duff:1983ajq} (for a review, see  \cite{Duff:1986hr}) 
but has regained some interest recently.
This is partly due to the swampland program (see, e.g.,  \cite{Vafa:2005ui}) and in particular to one of the conjectures proposed in this context, namely the 
{\it $AdS$ stability conjecture} \cite{Ooguri:2016pdq}: Any non-supersymmetric compactification leading to an $AdS$ spacetime  will be unstable.

Our universe is strongly believed, from observations, to be of de Sitter type for which another set of swampland conjectures indicate
that such compactifications  can never be stable, something that again may be due to the lack of supersymmetry in de Sitter space.
For these reasons it is of utmost important to prove these conjectures which so far have met with huge challenges. A second best approach is to
find examples supporting  their correctness or to search for counterexamples showing that the conjectures fail to be true (at least in their current formulation).

In the case of the AdS conjecture mentioned above a huge number of non-supersymmetric examples exist that are BF \cite{Breitenlohner:1982jf} (see below) stable
but which ultimately have been demonstrated to be unstable due to various kinds of decay channels. We will not discuss these any further here but instead
concentrate on the few cases that so far have not been proven unstable. Without claiming to be exhaustive we mentioned here two cases in this category:\\
\\
1. Certain S-folds \cite{Giambrone:2021wsm} and \\
2. Right-squashed $S^7$ \cite{ Duff:1983ajq}\\
\\
Note that without supersymmetry it seems very hard to prove stability directly but there are interesting attempts involving fake supersymmetry, see
\cite{Giri:2021eob}. The rest of this brief review is devoted to explaining recent results \cite{Nilsson:2018lof, Ekhammar:2021gsg, Nilsson:2023ctq, Karlsson:2023dnl} concerning the second example, the right-squashed $S^7$ compactification of D=11 supergravity.
Of course, it is not possible to claim stability just by testing known decay modes since there may be new ones found in the future. This is why supersymmetry, or possibly fake supersymmetry, 
is so utterly important for  proving stability.

When compactifying D=11 supergravity on $AdS_4\times S^7$ there are three cases to consider, the round case with eight supersymmetries, the left-squashed with one and the right-squashed case
with no supersymmetry. The two squashed cases are the skew-whiffed (orientation flipped) versions of each other.  Being supersymmetric the left-squashed case is absolutely stable from which
one can  show that also the right-squashed case is BF stable \cite{Breitenlohner:1982jf} since all unitarity bounds are respected also in the non-supersymmetric case \cite{Duff:1984sv}.

Going beyond  BF stability one may ask  if introducing interactions may ruin stability in the non-supersymmetric case. There are many aspects of this issue. One such is  addressed in \cite{Ahn:1999dq, Murugan:2016aty}, the up-shot of which  
is that marginal operators in the 1/N expansion of beta-functions in the boundary field theory are dangerous objects that might lead to the removal of fix-points and hence instabilities in the bulk theory.
It should be mentioned that the relevance of these claims are put in question by some authors, see, e.g., the Introduction in \cite{Karlsson:2023dnl} and work cited there. Here we will only discuss the presence 
or not of marginal operators which, as we will see, depends on our choice of boundary conditions.

To find the background solutions we need the bosonic part  of D=11 supergravity:
\begin{equation}
     \mathcal{L}
    = \frac{1}{ \kappa^2}\bigl(R - \frac{1}{12} F_{MNPQ} F^{MNPQ}
    + \frac{8}{12^4} \epsilon_{M_1 \hdots M_{11}} A^{M_1 M_2 M_3} F^{M_4 \hdots M_7} F^{M_8 \hdots M_{11}}\bigr).
\end{equation}
The ansatz for the D=11 background is given in terms of an 11= 4+7 split $X^M=(x^{\mu},\,y^m)$, a diagonal product metric  and a spacetime volume form for the 4-form field strength:
\beq
\bar G_{MN}=\text{diag}(\bar g_{\mu\nu}(x), \bar g_{mn}(y)),\,\,\bar F_{\mu\nu\rho\sigma}(x)=3m\bar\epsilon_{\mu\nu\rho\sigma}(x),
\eeq
where $m$ is a positive constant parameter of dimension $1/L$.

Inserting this ansatz into the field equations and Bianchi identities one finds the  background conditions
\beq
\bar R_{\mu\nu}=-12m^2\bar g_{\mu\nu},\,\,\,\bar R_{mn}(y)=6m^2\bar g_{mn}.
\eeq
Thus the background is $AdS_4$ times a seven-dimensional compact manifold $K^7$, examples of which are plentiful, see ,e.g., \cite{Duff:1986hr}.
As mentioned above we will here only discuss $K^7=S^7$. However, as explained in full detail in, e.g.,  \cite{Duff:1986hr} there exist one round and one squashed Einstein metric on
this manifold. While the orientation in the round case is irrelevant for the theory in $AdS_4$ this is not the case for the squashed metric as will be explained below.

One aim of this presentation is to summarize the recent results on the squashed  $S^7$ spectrum analysis, in particular the one for 2-forms,  
which finally has lead to an almost\footnote{This caveat is related to a degenaracy in the spectrum discovered in \cite{Karlsson:2023dnl}. See below.} complete understanding of all the fine details.
We start below by a short review of the irreps  relevant in $AdS_4$, i.e., the unitary  irreps of its isometry group $SO(2,3)$.

The author is very grateful  to  M.J.  Duff, C.N. Pope,  A. Padellaro, S. Ekhammar, and J. Karlsson for  the collaborations leading to the results presented here.
\section{SO(2,3) irrep diagrams  and singletons}

The unitary irreps of $SO(2,3)$, which is the isometry group of $AdS_4$ and also the conformal group on the boundary of $AdS_4$, are denoted $D(E_0, s)$ where 
$E_0$ is the energy of the lowest state (see the state diagram below copied from Nicolai \cite{Nicolai:1984hb}) and  $s$ its spin. In the case  of a massless (pseudo)scalar  ($s=0$) field 
$E_0$ is equal to 1 or 2 depending on the boundary conditions chosen for the field. Note that the diagram 
extends indefinitely upwards.

An interesting point for this presentation is the fact that the  irreps sitting at the unitarity bound values, $D(E_0=\half, s=0)$ and $D(E_0=1, s=\half)$ are known as
singletons \cite{Dirac:1963ta}. These have the special property that they fluctuate (or "live") only on the boundary of the $AdS_4$ 
space\footnote{That they can be gauged away in the bulk is discussed in  \cite{Casher:1984ym, Sezgin:2020avr}.} 
 \cite{Nicolai:1984gb, Samtleben:2024zoy}, 
being in some sense topological as bulk fields.
These irreps have state diagrams with states only along the main trajectory (the first lower right  one in the diagram below). 
Arguments  for why various kinds of   singletons are needed in order to
make sense of the different  $S^7$ spectra and their relations are presented for the first time in \cite{Nilsson:2018lof}, see also \cite{Nilsson:2023ctq, Karlsson:2023dnl}, and the discussion below
where a new aspect is mentioned\footnote{An early suggestion that singletons could be part of the round $S^7$ spectrum appears in \cite{Sezgin:1983ik}.}.

In some cases it is possible to follow the change in $E_0$, or rather the operator eigenvalues \cite{Nilsson:1983ru}, see also \cite{Nilsson:2023ctq, Karlsson:2023dnl}, when squashing which indicates that the singletons on the round $S^7$ must become ordinary bulk scalars and fermions in the squashed vacua  \cite{Karlsson:2023dnl} (see also \cite{Nilsson:2018lof, Nilsson:2023ctq}).
For this to be possible it seems that one must add to the singleton state diagrams states corresponding to bulk scalar or fermion fields which, if true, would be a new kind of Higgs effect. 
We will return to this issue again below. Unfortunately,
so far there is no Lagrangian realization of this in the literature as far as we know.

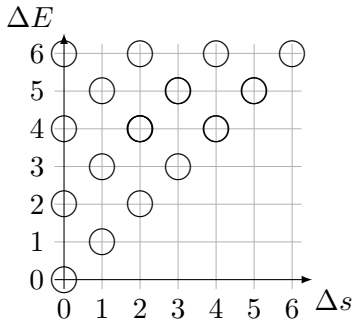
\begin{figure}[H]
\centering
\begingroup
\renewcommand{\scale}{0.5cm} 
\pgfmathsetmacro{\xrange}{6} 
\pgfmathsetmacro{\yrange}{6} 
\pgfmathsetmacro{\gridgreylevel}{30} 
\pgfmathsetmacro{\margin}{0.25}
\begin{tikzpicture}[x=\scale, y=\scale]
  \foreach \i in {1,...,\yrange} {
      \draw[very thin,black!\gridgreylevel] (-\margin,\i) node[left,black] {$\i$} -- ({\xrange+\margin},\i);
    }
  \foreach \i in {1,...,\xrange} {
      \draw[ultra thin,black!\gridgreylevel] (\i,-\margin) node[below,black] {$\i$} -- (\i,{\yrange+\margin});
    }

  \draw[-latex,black] (-\margin,0) node[left] {$0$} -- ([xshift=4pt]{\xrange+\margin},0) node[below,xshift=8pt] {$\Delta s$};
  \draw[-latex,black] (0,-\margin) node[below] {$0$} -- ([yshift=4pt]0,{\yrange+\margin}) node [left,yshift=6pt] {$\Delta E$};

  
    \node at (0,0) {$\bigcirc$};
      \node at (0,2) {$\bigcirc$};
        \node at (1,1) {$\bigcirc$};
          \node at (1,3) {$\bigcirc$};bigcirc
           \node at (0,4) {$\bigcirc$};
      \node at (0,6) {$\bigcirc$};
        \node at (1,5) {$\bigcirc$};
          \node at (3,5) {$\bigcirc$};
           \node at (2,2) {$\bigcirc$};
      \node at (2,4) {$\bigcirc$};
        \node at (3,3) {$\bigcirc$};
          \node at (3,5) {$\bigcirc$};
            \node at (2,4) {$\bigcirc$};
        \node at (4,4) {$\bigcirc$};
          \node at (5,5) {$\bigcirc$};
            \node at (2,6) {$\bigcirc$};
              \node at (4,6) {$\bigcirc$};
             \node at (6,6) {$\bigcirc$};
            \node at (2,4) {$\bigcirc$};
        \node at (4,4) {$\bigcirc$};
          \node at (5,5) {$\bigcirc$};

%


\end{tikzpicture}
\endgroup
\caption{Example of a state diagram for unitary irreps of $SO(2,3)$ denoted $D(E_0,s)$: For  massless (pseudo)scalars the states have energy and spin $(E,j)$ with $ E=E_0 + \Delta E$ where $E_0=1$ or $2$
(depending on boundary conditions)  and $j= s+\Delta s$ for $s=0$. The scalar singleton has $E_0=\frac{1}{2}$ and a state diagram consisting of only the main trajectory (the first  lower right one above). 
}
\end{figure}


\section{$D(E_0,s)$ and the relation of $E_0$ to the $S^7$ operator eigenvalues via the mass matrix }

In a Lagrangian formulation of a field theory on $AdS_4$ there are relations between the  $E_0$ of the irreps $D(E_0,s)$  and the masses
of the fields. These are given by the mass matrix relations in the table below where also the unitarity bound is given for each spin.


\begin{table}[H]
    \centering
    \renewcommand*{\arraystretch}{1.5} %
    \begin{tabular}{ll}
        \toprule
        $s=2$ & $E_0 = \frac{3}{2} + \frac{1}{2} {\displaystyle\sqrt{(M/m)^2+9}} \geq 3$ \\
        $s=\frac{3}{2}$ & $E_0 = \frac{3}{2} +\frac{1}{2}|{M/m-2}|\geq \frac{5}{2}$ \\
        $s=1$ & $E_0 = \frac{3}{2} + \frac{1}{2} {\displaystyle\sqrt{(M/m)^2 +1 }} \geq 2$ \\
        $s=\frac{1}{2}$ & $E_0 = \frac{3}{2} \pm \frac{1}{2}|{M/m}|\geq 1$ \\
        $s=0$ & $E_0 = \frac{3}{2} \pm \frac{1}{2} {\displaystyle\sqrt{(M/m)^2 +1}} \geq \frac{1}{2}$
        \\ \bottomrule
    \end{tabular}
    \caption{$E_0$ for $AdS_4$ fields of
    given mass $M$ and spin $s$ (in $Spin(2,3)$-irreps $D(E_0,s)$) and the corresponding unitarity bounds,
     see, e.g., \cite{Duff:1986hr}.}
    \label{ezerolist}
\end{table}

From  the full $D=11$ supergravity Lagrangian  linearized around any Freund-Rubin solution $AdS_4 \times K^7$ one can derive the relations between  
the mass matrices and operators on the compact internal manifold $K^7$. The idea is simply a generalization of the fact that in a flat spacetime $M_{11}=Mink_{4}\times T^7$ 
a $D=11$ box defines a mass matrix $M^2$ in $D=4$:
$\Box_{11}=\Box_4 + \Box_7 \equiv  \Box_4 - M^2$. The operators that appear this way on $K^7$ are $\Delta_p=\delta d + d \delta$ on $p$-forms for $p=0, 1, 2, 3$, the Lichnerowicz 
operator $\Delta_L$,  the linear operators $i\slashed{D}_{s}$ for $s=1/2, 3/2$. There is also the linear operator $Q=\star d$ acting on 3-forms related to $\Delta_3$ by $\Delta_3=Q^2$. 
The branches in the $0^+$ and $1^-$ spectra are due to the appearance of two fields of each kind in the diagonalization of the mass matrices in the analysis of $D=11$ supergravity.\\


\begin{table}[H]
    \centering
    \renewcommand*{\arraystretch}{1.2} %
    \begin{tabular}{ll}
        \toprule
        $\text{spin}^{\text{parity}}$ & Mass operators (giving $M^2$ for bosons and $M$ for fermions)
        \\ \midrule
        $2^{{+}}$ & $\Delta_0$ \\
        $\tfrac{3}{2}_{\scriptscriptstyle (\pm)}$ & $-i\slashed{D}_{1/2} + \tfrac{7m}{2}$ \\
        $1^{-}_{\scriptscriptstyle (\pm)}$ & $\Delta_1 + 12m^2 \pm 6m\sqrt{\Delta_1 + 4m^2}=(\sqrt{\Delta_1 + 4m^2}\pm3m)^2-m^2$ \\
        $1^+$ & $\Delta_2$ \\
        $\tfrac{1}{2}_{\scriptscriptstyle (\pm)}$ & $-i\slashed{D}_{1/2} - \tfrac{9m}{2}$ \\
        $\tfrac{1}{2}_{\scriptscriptstyle (\pm)}$ & $i\slashed{D}_{3/2} + \tfrac{3m}{2}$ \\
        $0^{+}_{\scriptscriptstyle (\pm)}$ & $\Delta_0 + 44m^2 \pm 12m\sqrt{\Delta_0 + 9m^2}=(\sqrt{\Delta_0 + 9m^2}\pm6m)^2-m^2$ \\
        $0^{+}$ & $\Delta_{L} - 4m^2=(\Delta_{L} - 3m^2)-m^2$ \\
        $0^{-}_{\scriptscriptstyle (\pm)}$ & $Q^2 + 6mQ + 8m^2=(Q+3m)^2-m^2$
        \\ \bottomrule
    \end{tabular}
    \caption{Mass operators  in Freund--Rubin compactifications, see, e.g.,  \cite{Duff:1986hr}.
    For  spins with two tower assignments, the subscripts $(\pm)$, the plus and minus signs refer to  branches of the $M^2$ formulas or to the positive and negative parts of the spectrum for
    linear operators.  Note the change of notation relative  \cite{Duff:1986hr} where superscripts $(1),(2)$ etc.\ were used instead of the $(\pm)$ notation of this paper.
    }
    \label{table:massop}
\end{table}

\section{Strategy to get the spectra of  operators on the squashed $S^7$}

When we now turn to the squashed seven-sphere we need a realization of it and a method to compute the spectra of the various operators 
appearing in the mass matrices as explained above. 

There are several ways to define the geometry of the squashed Einstein metric. The metric can for instance be obtained as the distance sphere in the quaternionic projective space ${\bf H}P^2$
as explained in detail in \cite{Duff:1986hr}, but perhaps a more intuitive picture is provided by
 the Hopf fibration, or Kaluza-Klein form, 
\beq
ds^2_{Hopf}(S^7)=d\mu^2+\frac{1}{4}\sin^2\mu\,\,\Sigma_i^2+\lambda^2\,(\sigma_i-A_i)^2,
\eeq
where $\Sigma_i$ and $\sigma_i$ are two sets of left-invariant 1-forms satisfying the $SU(2)$ Lie algebra, while $A_i$ is the $SU(2)$ $k=1$ instanton gauge field on $S^4$.
From this description of the squashed $S^7$ we see that squashing refers to a change in the  size of the fibre $S^3$ relative the base $S^4$.
The fact is that there are only two Einstein metrics which arise for $\lambda=1$, the round case, and for $\lambda=1/\sqrt 5$ which is the squashed case.
Although these two representations are frequently used we will here instead utilize a structure constant based approach for $(Sp_2\times Sp_1^C)/(Sp_1^A\times Sp_1^{B+C})$ 
developed in \cite{Bais:1983wc}. The strategy  has then following steps:\\
\\
1. Derive the full  spectrum of isometry irreps on the squashed $S^7$: Done in \cite{Nilsson:2018lof}.\\
2. Derive all possible operator eigenvalues: Done in \cite{Ekhammar:2021gsg, Karlsson:2021oxd, Karlsson:2023dnl}.\\
3. Tie the irreps in 1. to the eigenvalues in 2.: Done in \cite{Karlsson:2023dnl}, see also  \cite{Nilsson:2023ctq}.\\
4. Derive the possible values of $E_0$ for all fields in $AdS_4$  and form Heidenreich $\mN=1$ supermultiplets \cite{Heidenreich:1982rz}: Done in \cite{Karlsson:2023dnl}, see also  \cite{Nilsson:2023ctq}.\\
5. Skew-whiff  to the right-squashed non-supersymmetric squashed $S^7$ and study the possible single- and multi-trace operators on the $AdS_4$ boundary, and identify the set of marginal ones:
 Done in \cite{Karlsson:2023dnl}, see also  \cite{Nilsson:2023ctq}.
 
 A different approach used to derive parts of the squashed spectrum in $AdS_4$ was introduced in \cite{Duboeuf:2022mam} giving  results consistent with ours.


\section{Explicit 2-form mode results and their implications}

We start by explaining the method used to obtain the isometry irrep content of the squashed $S^7$ compactification. To be concrete and simple
we start by discussing how to get the spectrum of spin two modes in $AdS_4$. Recall that the $SO(2,3)$ irreps are denoted $D(E_0, s)$ where for $s=2$
we have $E_0=\frac{3}{2}+\frac{1}{6}\sqrt{M^2/m^2+9}$ where the mass matrix $M^2=\Delta_0$. Thus to find the spin 2 spectrum in $AdS_4$ we need to derive the 
eigenvalue spectrum of $\Delta_0=-\Box$ on scalar modes on the squashed $S^7$ and the spectrum of isometry irreps that can occur. To obtain the latter the general procedure 
for any tensor field on a coset manifold $G/H$ is:\\
\\
1. Split the tangent space irrep of the field into $H$ irreps,\\
2. The isometry spectrum contains any $G$ irrep that when decomposed under $H$ contains any of the $H$ irreps found in the previous step.\\
\\
 The result for the squashed $S^7$ is presented in terms of 
{\it cross diagrams} in \cite{Nilsson:2018lof} where each cross corresponds to an $G=Sp_2\times Sp_1^C$ irrep $(p,q;r)$. 
For scalars  $\Delta_0=\frac{20m^2}{9}C_G$, where $C_G(p,q;r)$ is the Casimir,
and the spectrum is given by the diagram (note that  $r=p$),
\begin{figure}[H]
    \centering
    \begingroup
    \renewcommand{\scale}{0.50cm}  %
    \footnotesize
    \newcommand{\fig}{%
        \begin{tikzpicture}[x=\scale, y=\scale]
            \crossdiagramlayout{p}
            \foreach \p in {0,...,\range} {
                \foreach \q in {0,...,\range} {
                    \node at (\p,\q) {$\times$};
                }
            }
        \end{tikzpicture}
    }
    \newcommand*{\eig}[1]{{\normalsize$\Delta_0^{#1}$}}%
    \begin{tabular}{c}
        \eig{(1)}   \\
        \fig
    \end{tabular}
    \endgroup
    \caption{Scalar cross diagram. Each cross corresponds to an isometry irrep $(p,q;r)$, for $p\ge 0$, $q\ge 0$, and $r=p$ of eigenmodes $\phi$ of $\Delta_0$ with eigenvalues
    $\Delta_0^{(1)} =\frac{m^2}{9} \,20C_g$, as given in \cite{Nilsson:2018lof}.
    }
\end{figure}
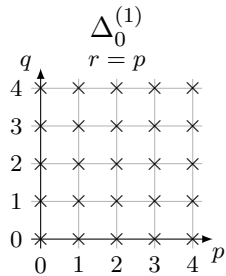

In the case of 2-forms \cite{Nilsson:2018lof} the isometry irrep spectrum contains 21 cross diagrams 15 of which contain transverse modes. 
The corresponding 15 cross diagrams are divided into sets that make up $Sp_1^C$ irreps, in fact we have
one  ${\bf 1}$, three ${\bf 3}$ and one ${\bf 5}$ \cite{Karlsson:2023dnl}. These appear as the vertical sets in the table below.

\begin{figure}[H]
\label{twoformcrosses}
  \centering
  \begingroup %
  \renewcommand{\scale}{0.41cm} %
  \footnotesize %
  \newcommand{\figIp}{%
      \begin{tikzpicture}[x=\scale, y=\scale]
          \crossdiagramlayout{p}
          \foreach \p in {1,...,\range} {
              \foreach \q in {0,...,\range} {
                  \node at (\p,\q) {$\times$};
              }
          }
          \foreach \q in {1,...,\range} {
              \node at (0,\q) {$\times$};
          }
      \end{tikzpicture}%
  }%
  \newcommand{\figIIppII}{%
      \begin{tikzpicture}[x=\scale, y=\scale]
          \crossdiagramlayout{p+2}
          \foreach \p in {1,...,\range} {
              \foreach \q in {0,...,\range} {
                  \node at (\p,\q) {$\times$};
              }
          }
          \foreach \q in {2,...,\range} {
              \node at (0,\q) {$\times$};
          }
      \end{tikzpicture}%
  }%
  \newcommand{\figIIp}{%
      \begin{tikzpicture}[x=\scale, y=\scale]
          \crossdiagramlayout{p}
          \foreach \p in {2,...,\range} {
              \foreach \q in {0,...,\range} {
                  \node at (\p,\q) {$\times$};
              }
          }
          \foreach \q in {1,...,\range} {
              \node at (1,\q) {$\times$};
          }
      \end{tikzpicture}%
  }%
  \newcommand{\figIIpmII}{%
      \begin{tikzpicture}[x=\scale, y=\scale]
          \crossdiagramlayout{p-2}
          \foreach \p in {3,...,\range} {
              \foreach \q in {0,...,\range} {
                  \node at (\p,\q) {$\times$};
              }
          }
          \foreach \q in {1,...,\range} {
              \node at (2,\q) {$\times$};
          }
      \end{tikzpicture}%
  }%
  \newcommand{\figIIIppII}{%
      \begin{tikzpicture}[x=\scale, y=\scale]
          \crossdiagramlayout{p+2}
          \foreach \p in {0,...,\range} {
              \foreach \q in {1,...,\range} {
                  \node at (\p,\q) {$\times$};
              }
          }
      \end{tikzpicture}%
  }%
  \newcommand{\figIIIp}{%
      \begin{tikzpicture}[x=\scale, y=\scale]
          \crossdiagramlayout{p}
          \foreach \p in {1,...,\range} {
              \foreach \q in {0,...,\range} {
                  \node at (\p,\q) {$\times$};
              }
          }
      \end{tikzpicture}%
  }%
  \newcommand{\figIIIpmII}{%
      \begin{tikzpicture}[x=\scale, y=\scale]
          \crossdiagramlayout{p-2}
          \foreach \p in {2,...,\range} {
              \foreach \q in {0,...,\range} {
                  \node at (\p,\q) {$\times$};
              }
          }
      \end{tikzpicture}%
  }%
  \newcommand{\figIVppII}{%
      \begin{tikzpicture}[x=\scale, y=\scale]
          \crossdiagramlayout{p+2}
          \foreach \p in {0,...,\range} {
              \foreach \q in {0,...,\range} {
                  \node at (\p,\q) {$\times$};
              }
          }
      \end{tikzpicture}%
  }%
  \newcommand{\figIVp}{%
      \begin{tikzpicture}[x=\scale, y=\scale]
          \crossdiagramlayout{p}
          \foreach \p in {1,...,\range} {
              \foreach \q in {1,...,\range} {
                  \node at (\p,\q) {$\times$};
              }
          }
      \end{tikzpicture}%
  }%
  \newcommand{\figIVpmII}{%
      \begin{tikzpicture}[x=\scale, y=\scale]
          \crossdiagramlayout{p-2}
          \foreach \p in {2,...,\range} {
              \foreach \q in {0,...,\range} {
                  \node at (\p,\q) {$\times$};
              }
          }
      \end{tikzpicture}%
  }%
  \newcommand{\figVppIV}{%
      \begin{tikzpicture}[x=\scale, y=\scale]
          \crossdiagramlayout{p+4}
          \foreach \p in {0,...,\range} {
              \foreach \q in {1,...,\range} {
                  \node at (\p,\q) {$\times$};
              }
          }
      \end{tikzpicture}%
  }%
  \newcommand{\figVppII}{%
      \begin{tikzpicture}[x=\scale, y=\scale]
          \crossdiagramlayout{p+2}
          \foreach \p in {1,...,\range} {
              \foreach \q in {2,...,\range} {
                  \node at (\p,\q) {$\times$};
              }
          }
      \end{tikzpicture}%
  }%
  \newcommand{\figVp}{%
      \begin{tikzpicture}[x=\scale, y=\scale]
          \crossdiagramlayout{p}
          \foreach \p in {2,...,\range} {
              \foreach \q in {1,...,\range} {
                  \node at (\p,\q) {$\times$};
              }
          }
      \end{tikzpicture}%
  }%
  \newcommand{\figVpmII}{%
      \begin{tikzpicture}[x=\scale, y=\scale]
          \crossdiagramlayout{p-2}
          \foreach \p in {3,...,\range} {
              \foreach \q in {1,...,\range} {
                  \node at (\p,\q) {$\times$};
              }
          }
      \end{tikzpicture}%
  }%
  \newcommand{\figVpmIV}{%
      \begin{tikzpicture}[x=\scale, y=\scale]
          \crossdiagramlayout{p-4}
          \foreach \p in {4,...,\range} {
              \foreach \q in {0,...,\range} {
                  \node at (\p,\q) {$\times$};
              }
          }
      \end{tikzpicture}%
  }%
  \newcommand*{\eig}[1]{{\normalsize$\Delta_{2}^{#1}$}}%
  \begin{tabular}{ccccc}%
                &               &               &                 & \eig{(3)'}    \\
                & \eig{(2)_-}   & \eig{(2)_+}   & \eig{(3)}       & \figVppIV     \\
      \eig{(1)} & \figIIIppII   & \figIVppII    & \figIIppII      & \figVppII     \\
      \figIp    & \figIIIp      & \figIVp       & \figIIp         & \figVp        \\
                & \figIIIpmII   & \figIVpmII    & \figIIpmII      & \figVpmII     \\
                &               &               &                 & \figVpmIV
  \end{tabular}%
  \endgroup
  \caption{Transverse two-form cross diagrams. 
  }
  %
\end{figure}
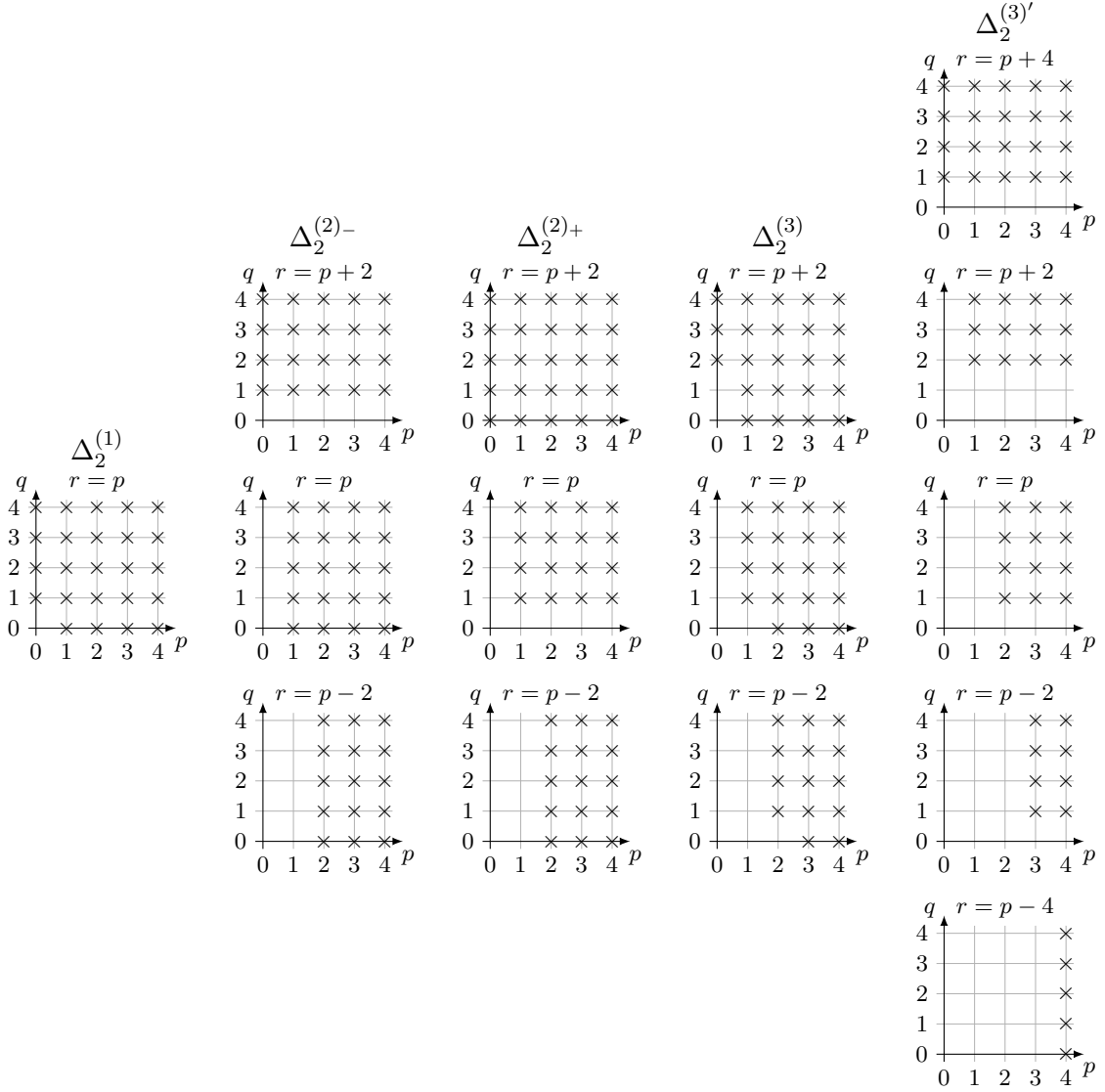

By inspection we see that some diagrams lack  crosses along some lines 
parallell to the $p$-axes which is due to these irreps having zero norm \cite{Karlsson:2023dnl}.
This phenomenon was mentioned for spin 1/2 modes already in \cite{Nilsson:1983ru}. 

The eigenvalues for these transverse 2-form modes were first calculated 
in \cite{Ekhammar:2021gsg} using a method relying heavily on $G_2$ and octonions, but which did not tie the eigenvalues to specific crosses in the table above. 
An improved version of this method was later developed in \cite{Karlsson:2021oxd, Karlsson:2023dnl}
giving rise to  the following novel formulas: The algebraic approach to a (reductive) coset $G/H$ gives a relation between a tangent space differential operator and a Lie algebra element $T_a$ in the coset, namely
$\check{D}_a=D_a+\half f_{abc}\Sigma^{bc} =-T_a$  where the structure constants are defined by $[T_a, T_b]=f_{ab}{}^cT_c+ f_{ab}{}^iT_i$ and $\Sigma^{ab}$ are the generators of the tangent space group. 
For the squashed $S^7$ it was pointed out in \cite{Bais:1983wc} that  $f_{abc}=-\frac{1}{\sqrt 5}a_{abc}$,  where $a_{abc}$ are the octonionic structure constants,  implying that the H-covariant  dserivative $\check{D}_a=D_a-\frac{1}{2\sqrt 5}a_{abc}\Sigma^{bc} $
is also 
 $G_2$ covariant in the sense that $\check{D}_a a_{abc}=0$ \cite{Ekhammar:2021gsg}.
Using  these properties a universal Laplacian $\Delta=-\Box-R_{abcd}\Sigma^{ab}\Sigma^{cd}$  (unifying $\Delta_p$ and $\Delta_L$) was found in \cite{Karlsson:2021oxd} and used 
heavily in \cite{Karlsson:2023dnl}. It leads to the following group theoretic version of the  operator equation on the squashed $S^7$ valid for any of the relevant tangent space tensors \cite{Karlsson:2021oxd}: 
\beq
\Delta=C_g+\frac{6}{7}C_{SO(7)}-\frac{3}{2}C_{G_2}-\frac{1}{\sqrt 5}a_{abc}\Sigma^{ab}\check{D}^c,
\eeq 
where the $C$s are Casimir operators and $\Sigma^{ab}$ are the $SO(7)$ generators. Note that projectors onto $G_2$ irreps are needed when solving the eigenvalue equations involving this $\Delta$  
 and that these projectors are all written in terms of
octonionic structure constants \cite{Karlsson:2021oxd, Karlsson:2023dnl}.

So far we have not been able to associate the eigenvalues to the isometry irreps in the cross diagrams. 
To address this mode-eigenvalue issue we need explicit expressions for the mode functions: The  operators below acting on scalar modes give all  21 2-form modes.
  Note  that they contain the octonionic structure constants
  $a_{abc}$ and its dual $c_{abcd}$  as well as the $Sp_1^C$ Killing vectors $s^i=s^{ia}\partial_a$.

 \beqa
  { \mathcal Y}^{(1)}_{ab} &=& a_{ab}{}^c\check{D}_c, \,\,\,\,\,
       { \mathcal Y}^{(2)i}_{ab} = a_{ab}{}^c s_c{}^i,\,\,\,\,\,
      { \mathcal Y}^{(3)i}_{ab} = \epsilon^i{}_{jk} s_a{}^j s_b{}^k,\,\,\,\,
       { \mathcal Y}^{(4)i}_{ab} = s_{[a}{}^i \check{D}_{b]},\\
        { \mathcal Y}^{(5)i}_{ab} &=& c_{ab}{}^{cd} s_c{}^i \check{D}_d,\,\,\,\,
        { \mathcal Y}^{(6)i}_{ab} = a_{[a|}{}^{cd} s_{c}{}^i \check{D}_{d|b]},\,\,\,\,
   { \mathcal Y}^{(7)ij}_{ab} = s_{[a}{}^{\{i|} a_{b]}{}^{cd} s_c{}^{|j\}} \check{D}_d.
\eeqa
The novel aspect of these mode functions is present in $ { \mathcal Y}^{(6)i}_{ab}$: It contains a two derivative  operator $\check{D}_{ab}=\check{D}_{(a}\check{D}_{b)}$.
 This was found in  \cite{Karlsson:2023dnl} and will be further
explained  in \cite{Karlsson:2024xxx}. Computing the eigenvalues by acting with the universal Laplacian above directly on the mode functions does of course provide the connection
between eigenvalues and Fourier modes (crosses in the cross diagram). These results confirm that there is a degenaracy in the spectrum since it is found that
$\Delta_2^{(3)}=\Delta_2^{(3)'}$ in Table \ref{twoformcrosses} above \cite{Karlsson:2023dnl}.

When the entire squashed spectrum was derived in \cite{Nilsson:2018lof} it was also found that all states fit into the following $\mN=1$ supermultiplets (referred to by their maximum spin component):
1 spin 2, 6 spin 3/2, 6 spin $1^-$, 8 spin $1^+$ and 14 Wess-Zumino multiplets. This was verified in \cite{Karlsson:2023dnl} by demonstrating that the $E_0$ values fit this structure.
%
However, when comparing this squashed  spectrum to the round one some states do not fit into a Higgs picture relating the two spectra. These states are of two types \cite{Nilsson:2018lof}:\\
\\
1. States occurring in the round spectrum when breaking $SO(8) \rightarrow Sp_2 \times Sp_1^C$ but not in the squashed one: \\
Lichnerowicz modes  of $\Delta_L:$\,\,\,$(\bf 4, \bf 2), (\bf 5, \bf 3)$\\
 Rarita-Schwinger modes of $i\slashed{D}_{3/2}:$\,\,\, $(\bf 4, \bf 2),  (\bf 5, \bf 1), (\bf 1, \bf 3)$\\
2. A state occurring in the squashed spectrum but not in the round: $i\slashed{D}_{3/2}$ mode $({\bf 1}, {\bf 1})$ \cite{Nilsson:2023ctq}.\\
\\
A way to make these facts fit with a Higgs/deHiggs kind of picture was suggested in \cite{Nilsson:2018lof} (see also \cite{Nilsson:2023ctq, Karlsson:2023dnl}) 
 involving adding singletons to the spectra: An $\mN=8$ supersingleton to the round spectrum and 
a fermionic singlet singleton to right-squashed spectrum. One strong indication that adding the singletons is really necessary is that one can follow the  round sphere fermionic singleton 
in the $(\bf 1, \bf 3)$ irrep  when squashing,
and find that it appears as the superpartner of the gauge field  with $E_0=3/2$ in the squashed vacuum \cite{Karlsson:2023dnl}\footnote{This can be done also for the other two pieces of the supersingleton.}. 
But then the singleton state 
diagram must have picked up a bulk worth of fermionic states  to become a proper bulk field.
 In the squashed vacuum these new states must have  $E_0=5/2$ as can be seen from  Nicolai's state diagram in  Figure 2  of  \cite{Nicolai:1984hb}.   This value  should match the $E_0$ obtained by following,
 if possible, the corresponding Rarita-Schwinger mode in this irrep from the round to the squashed vacuum.  This is not yet possible but from the Rarita-Schwinger results in \cite{Karlsson:2023dnl} 
 (see Eqs. (3.100) - (3.103))
 one can check where the 
 $(\bf 1, \bf 3)$  mode would have ended up in the squashed vacuum and confirm that $E_0=5/2$ is obtained. A singleton Higgs effect of this kind could have its origin in the Flato-Fronsdal dipole discussed in \cite{Flato:1986uh}.

\begin{table}[ht]
    \centering
    \setlength{\defaultaddspace}{4pt}
    \begin{tabular}{lllllcl}
        \toprule
        $s^p$ & $E_0$ & $E_0-\frac{1}{2}$ & $E_0+\frac{1}{2}$ & $E_0$  & $Sp_1^C$ & $E_0$ values\\ \midrule
        $2^+$ & $2^+(\Delta_0^{(1)})$ & $\frac{3}{2}(i\slashed{D}_{1/2}^{(1)_+})$ & $\frac{3}{2}(i\slashed{D}_{1/2}^{(1)_-})$ &$1^+(\Delta_2^{(1)})$ & $\bf{1}$ & $\frac{3}{2}+\frac{1}{6}\sqrt{20C_g+81}$\\ \addlinespace
        $\frac{3}{2}_1$ & $\frac{3}{2}(i\slashed{D}_{1/2}^{(2)_-})$ & $1^+(\Delta_2^{(2)_+})$ &$ 1_{\scriptscriptstyle (+)}^-(\Delta_1^{(1)_-}) $& $\frac{1}{2}(i\slashed{D}_{3/2}^{(3)_+})$ &$\bf{3}$ &  $\frac{3}{2}+\frac{5}{6}+\frac{1}{6}\sqrt{20C_g+49}$\\ \addlinespace
        $\frac{3}{2}_2$ & $\frac{3}{2}(i\slashed{D}_{1/2}^{(2)_+})$ &  $1_{\scriptscriptstyle (-)}^-(\Delta_1^{(1)_+}) $ & $1^+(\Delta_2^{(2)_-})$ & $\frac{1}{2}(i\slashed{D}_{3/2}^{(3)_-})$ &$\bf{3}$  &  $\frac{3}{2}-\frac{5}{6}+\frac{1}{6}\sqrt{20C_g+49}$\\ \addlinespace
        $1^-_1$ & $1_{\scriptscriptstyle (+)}^-(\Delta_1^{(1)_+})$ & $\frac{1}{2}(i\slashed{D}_{3/2}^{(4)_+})$ & $\frac{1}{2}(i\slashed{D}_{1/2}^{(2)_+})$ & $0^-(Q^{(2)_+})$&$\bf{3}$  &  $\frac{3}{2}+\frac{5}{3}+\frac{1}{6}\sqrt{20C_g+49}$\\ \addlinespace
        $1^-_2$ & $1_{\scriptscriptstyle (-)}^-(\Delta_1^{(1)_-})$ & $\frac{1}{2}(i\slashed{D}_{1/2}^{(2)_-})$ & $\frac{1}{2}(i\slashed{D}_{3/2}^{(4)_-})$ & $0^-(Q^{(2)_-})$&$\bf{3}$  &  $\frac{3}{2}-\frac{5}{3}+\frac{1}{6}\sqrt{20C_g+49}$\\ \addlinespace
        $1^+_1$ & $1^+(\Delta_2^{(3)})$ &  $\frac{1}{2}(i\slashed{D}_{3/2}^{(2)_-})$ &  $\frac{1}{2}(i\slashed{D}_{3/2}^{(2)_+})$ &$0^+(\Delta_L^{(1)})$ &$\bf{3}$  &  $\frac{3}{2}+\frac{1}{6}\sqrt{20C_g+9}$\\ \addlinespace
        $1^+_2$ & $1^+(\Delta_2^{(3)'})$ &  $\frac{1}{2}(i\slashed{D}_{3/2}^{(2)'_-})$ & $\frac{1}{2}(i\slashed{D}_{3/2}^{(2)'_+})$ &$0^+(\Delta_L^{(1)'})$ &$\bf{5}$  &  $\frac{3}{2}+\frac{1}{6}\sqrt{20C_g+9}$
        \\ \bottomrule
    \end{tabular}
    \caption{Supermultiplets with spin  (and parity) $s=1^{\pm}$, $s=3/2$, and  $s=2$  \cite{Karlsson:2023dnl}. Here, each entry,  represented by  $s(\text{operator}_{\text{mode type}}^{\text{eigenvalue}})$,
    corresponds to a specific  spin component of a Heidenreich $\mN=1$ supermultiplet, but with the highest spin first.
    The notation indicates also the relevant  cross diagrams of $\Delta_p$ (for $p=0,1, 2, L$), $Q$, $i\slashed{D}_{1/2}$ , and $i\slashed{D}_{3/2}$ as given in  \cite{Karlsson:2023dnl}.
    The $Sp_1^C$ irrep entries specify the number of cross diagrams belonging to the supermultiplet. 
    The short, i.e., massless, supermultiplets are $2^+$ for $(p,q;r)=(0,0;0)$ and $1^-_2$ for $(2,0;0)$ and $(0,0;2)$.
    }
    %
\end{table}

Note that in the last two lines of the above table the degenerate eigenvalues $\Delta_2^{(3)}=\Delta_2^{(3)'}$ occur but
 that the corresponding eigenfunctions are different (and independent) as is clear from the previous discussion. This is of course also the case for all the other components 
 of the last two supermultiplets in the table above. 
One should also note how the above 15 cross diagrams for $\Delta_2$  are spread out over the whole supermultiplet table.

For the Wess-Zumino supermultiplets in the table below another interesting phenomenon appears: Supersymmetry does not uniquely specify the supermultiplet structure
since, for the $WZ_2$ case, one can by choosing boundary conditions find  two ways to satisfy supersymmetry. This is connected to $\pm$ sign appearing in the formula for the energy eigenvalues given in the table
for $WZ_2$,  namely $E_0=\frac{3}{2}\pm \bigl(-\frac{5}{2}+\frac{1}{6}\sqrt{20C_g+81}\bigr)$. This table  contains another case of degeneracy, this time without involving $\Delta_2$, as is seen 
in the last four lines. In all the degenerate cases $\Delta_L$ occurs which will be further discussed in \cite{Karlsson:2024xxx}.

The final remaining supermultiplets are the Wess-Zumino ones with $C_g=0$, i.e., the singlets. The entire  set of singlet fields in $AdS_4$ along with their explicit mode functions on the squashed $S^7$ was analysed in detail in \cite{Nilsson:2023ctq}, both for left and right squahing. 

\begin{table}[H]
    \centering
    \setlength{\defaultaddspace}{4pt}
    \begin{tabular}{lll @{$\mskip\thinmuskip$}c@{$\mskip\thickmuskip$} lcl}
        \toprule
        Mult. & $E_0$ &$E_0-\frac{1}{2}$ && $E_0+\frac{1}{2}$ &   ${Sp_1^C}$& $E_0$ values\\ \midrule
        WZ1 &$\frac{1}{2}(i\slashed{D}^{(1)_+}_{1/2})$ & $0^-(Q^{(3)_+})$ && $0_{\scriptscriptstyle (+)}^+(\Delta_0^{(1)})$ & $\bf{1}$ & $\frac{3}{2}+\frac{5}{2}+\frac{1}{6}\sqrt{20C_g+81}$\\ \addlinespace
        WZ2 & $\frac{1}{2}(i\slashed{D}^{(1)_-}_{1/2})$ & $0^+_{\scriptscriptstyle (-)}(\Delta_0^{(1)})$ & $\leftrightarrow$ & $0^-(Q^{(3)_-})$ & $\bf{1}$ & $\frac{3}{2}\pm \bigl(-\frac{5}{2}+\frac{1}{6}\sqrt{20C_g+81}\bigr)$\\ \addlinespace
        WZ3 &$\frac{1}{2}(i\slashed{D}_{3/2}^{(1)_+})$ & $0^+(\Delta_L^{(2)_+})$ && $0^-(Q^{(1)_+})$  & $\bf{1}$  & $\frac{3}{2}+\frac{5}{6}+\frac{1}{6}\sqrt{20C_g+1}$\\ \addlinespace
        WZ4 &$\frac{1}{2}(i\slashed{D}_{3/2}^{(1)'_+})$ & $0^+(\Delta_L^{(2)'_+})$ && $0^-(Q^{(1)'_+})$  & $\bf{5}$  & $\frac{3}{2}+\frac{5}{6}+\frac{1}{6}\sqrt{20C_g+1}$\\ \addlinespace
        WZ5 & $\frac{1}{2}(i\slashed{D}_{3/2}^{(1)_-})$ &$0^-(Q^{(1)_-})$ && $0^+(\Delta_L^{(2)_-})$ & $\bf{1}$  &  $\frac{3}{2}-\frac{5}{6}+\frac{1}{6}\sqrt{20C_g+1}$\\ \addlinespace
        WZ6 & $\frac{1}{2}(i\slashed{D}_{3/2}^{(1)'_-})$ &$0^-(Q^{(1)'_-})$ && $0^+(\Delta_L^{(2)'_-})$ & $\bf{5}$  &  $\frac{3}{2}-\frac{5}{6}+\frac{1}{6}\sqrt{20C_g+1}$
        \\ \bottomrule
    \end{tabular}
    \caption{Wess--Zumino supermultiplets for $C_g>0$. See \cite{Karlsson:2023dnl} for definitions.
    The upper sign in the WZ2 $E_0$ value corresponds to $(D,D,D)$ boundary conditions for $p+q \geq 3$, $(D,N,D)$ for $p+q=2$ and $(N,N,D)$ for $p+q=1$.
    The scalar $0^+$ and pseudo-scalar $0^-$ should change places in the WZ2 supermultiplet if the lower sign in the $E_0$ formula is used, as indicated by the arrow.
    The lower sign is only valid for $p+q = 2$ and $p+q = 1$ and then corresponds to $(N,N,D)$ and $(D,N,D)$ boundary conditions, respectively.
    The subscript on $0^+_{\scriptscriptstyle (\pm)}$
    indicates which branch of $M^2(\Delta_0)$ is used.}
    \label{wesszuminopositivecgmultiplets}
\end{table}

\section{Conclusions}

We end by listing a number of conclusions related to the results presented above:\\
1. {\it Singletons}:  There are modes in the round case that do not appear in either of the squashed cases although the latter are a kind of Higgsed version of the former.
However, by adding the $\mN =8$ singleton supermultiplet to the round spectrum, and a fermionic singlet singleton to the right-squashed spectrum, 
 one can argue that the three spectra are consistent with each other (see, e.g., the discussion above Table 3).
There is also a spin 3/2 mode in the squashed spectrum that does not exist in the round spectrum. This fact can also be seen to be consistent
by invoking a deHiggsing in the spin 3/2  sector of the two squashed cases  \cite{Nilsson:2018lof, Nilsson:2023ctq}.\\
2. {\it The left-squashed spectrum 1}:  The $\mN =1$ supersymmetric spectrum in the left-squashed case is  completely understood apart from  the degeneracy encountered
in all supermultiplets containing fields whose masses are related to the Lichnerowicz operators on $S^7$ \cite{Karlsson:2023dnl}.\\
3.  {\it The left-squashed spectrum 2}:  We find a rather surprising feature namely that supersymmetry does not fix the boundary conditions of scalar and spin 1/2 fields in the left-squashed case.
In other words, the values of $E_0$ for these fields fit in two different ways into Wess-Zumino supermultiplets using different assignments of boundary conditions
(that is different choices of $\pm$ in the formulas for $E_0$). In fact, there are  irreps whose  boundary conditions in the round case cannot be retained in the squashed case \cite{Karlsson:2023dnl}
indicating that there is no general relation of this kind.\\
4.  {\it The right-squashed spectrum 1}:  By skew-whiffing (orientation-flipping) one obtains the right-squashed non-supersymmetric compactification which again is BF stable \cite{Duff:1984sv}. \\
5.  {\it The right-squashed spectrum 2}: Having established
in the left-squashed case that boundary conditions can be chosen in different ways, the right-squashed spectrum becomes very dependent on boundary conditions and how they 
are chosen. In particular, it is possible to choose boundary conditions  so that there are no marginal operators on the boundary at all  \cite{Nilsson:2023ctq, Karlsson:2023dnl}, something that might be 
relevant for the $AdS$ swampland stability issue \cite{Ooguri:2016pdq}. \\

\end{document}